\documentstyle[preprint,aps]{revtex}

\begin{document}
\def\be7pg{$^7Be(p,\gamma)^8B$}
\def\b8{$^8B$}
\draft
\title{Comment on E2 Contribution to the $^8B\ \rightarrow\ p\ +\ ^7Be$
Coulomb Dissociation Cross Section. [K. Langanke and T.D. Shoppa,
Phys. Rev. {\bf C49}, R1771(1994), ER(1994)]}

\author{Moshe Gai}
\address{Dept.of Physics, Yale University, New Haven, CT 06511}

\author{Carlos A. Bertulani}
\address{Gessellschaft f\"{u}r Schwerionenforchung, KPII,
Planckstr. 1, D-64291 Darmstadt, Germany}

\maketitle

\newpage

\begin{abstract}
The E2 cross section calculated by Langanke and Shoppa for the Riken
experiment on the Coulomb Dissociation of \b8 uses E2 nuclear
matrix element from one sepcific model.
Other nuclear models predict a consdiderably smaller E2 cross section
(by approximately a factor of 3 to 4), and
Langanke and Shoppa appear to assume the most
optimistic scenario predicting a large
E2 cross section.  We also note that Barker
has already criticised the nuclear
model used by Langanke and Shoppa.
A {\bf model independent} Chi-Square analysis
of the Riken data, suggest the best
fit for the current Riken data is obtained with E1 amplitudes only.
The upper limit (90\% confidence) on the E2 component
derived from our Chi-Square
analysis is considerably smaller than
used by Langanke and Shoppa. The model dependent analysis of
Langanke-Shoppa should not be considered as a correction to the Riken
result, as claimed, and their quoted $S_{17}(0)$ is not substantiated.

\end{abstract}

\pacs{PACS numbers: 25.20.+y, 25.70.+a, 27.20.+n, 96.20.Ka}

In a recent publication Langanke and Shoppa (LS)
\cite{Lang1,Lang2} calculated the E2 cross section
for the Riken experiment on the Coulomb Dissociation of \b8 \cite{Moto}.
The measured cross section of the
\be7pg reaction includes contributions from s and d -waves
($S_{E1}$) p-waves ($S_{M1}\ and\ S_{E2}$) and f-waves ($S_{E2}$), where
the p-wave cross section is dominated by a resonance at $E_{cm}=632\ keV$.
All these amplitudes contribute to the measured Coulomb Dissociation of
\b8,  with the E1 being dominant, and the (small) E2 enhanced due to the
large virtual photon flux (especially at large angles, $\theta\ \geq \
4 ^\circ$). In this comment on the work of LS we concentrate on the
E2 cross section of the Coulomb Dissociation of \b8 and ignore the M1
cross section, even though the M1 appears to contribute to the 600 keV
angular distribution data of Motobayashi
et al. \cite{Moto} at a level comparable to that
of the E2 (10\%), which is however smaller
then the quoted accuracy of the Riken experiment (15-20\%).

Recently one of us (MG) notified LS of a mistake in their original paper
\cite{Lang1} which led to the publication of an Erratum \cite{Lang2} with
a correction of Fig. 2 of LS \cite{Lang1}.  We however note that Fig. 1
of LS \cite{Lang1} is in fact still misleading as
it was not corrected and thus does not reflect the E2
contribution predicted for the Riken experiment.  In both
publications of LS they
quote a value for $S_{17}(0)$ (with one decimal point accuracy!) creating
the impression that they had carried out precission correction to the Riken
result.  In fact this value of $S_{17}(0)$
was quoted by Bahcall et al. \cite{Bah} that
support the notion that a substantial correction must be applied to
the Riken data, as they
state: "When the E2 contribution to this reaction is taken into account
\cite{Lang1}, the preliminary Coulomb-Dissociation value differs from the six
direct measurements of the \be7pg reaction by a factor of two...", contrary
to data shown by Motobayashi et al. \cite{Moto}. We also emphasize that while
the LS "correction factors" are deduced with large uncertainties,
$1\ -\ \alpha\ =\ 0.27\pm 0.1,\ 0.18\pm 0.16,\ and\ 0.23\pm 0.17$, at
0.6, 0.8, and 1.0 MeV, respectively, the
so called corrected $S_{17}(0)$ is quoted \cite{Lang2}
with the experimental error only ($\pm 3\ eV$) \cite{Moto}.

In this comment on LS work we demonstrate that
LS appear to have considered the most optimistic scenario for a large E2
contribution, which makes their {\bf model dependent} analysis
questionable.  We present here a preliminary {\bf model independent} analysis
that does not support LS conclusion and their quoted value of $S_{17}(0)$.

The construction of a reliable Nuclear Structure model for the \be7pg
reaction has received a great deal of theoretical attention
\cite{Kim,Bark1,Bark2,Baye,Krauss,Typel}. Once such a model is constructed
the predicted cross sections could be used in conjunction with
the formalism developed by Baur, Bertulani, and
Rebel \cite{Baur} to calculate (differential) cross sections for the Coulomb
Dissociation of \b8.  Current Nuclear Structure models are in agreement
(approximately 10-20\%) on the calculated E1 (and resonant M1) cross section
of the \be7pg reaction, but in
disagreement on the predicted small E2 cross section.  For example
while Kim et al. \cite{Kim} predict for the 632 keV resonance
$S_{E2}/S_{E1}\ =\ 1.8 \times 10^{-3}$, Typel and Baur \cite{Typel}
predict a value of $5 \times 10^{-4}$, almost a factor of 4 smaller.
While Krauss et al. \cite{Krauss} are in agreement with Kim et al.,
Descouvemont and Baye \cite{Baye} predict for the 632 keV resonance a
$B(E2:\ 2^+\ \rightarrow \ 1^+)\ =\ 13\ W.u.$ which is a factor of
2.4 smaller than predicted by Kim et al.  We note that Barker
has constructed a model that includes the 632 keV resonance which allows
for predicting all required cross sections including the E2 \cite{Bark1}.
Such calculations
yield E2 (and M1) cross sections which are considerably smaller
\cite{Bark1} than predicted by Kim et al. In fact Barker has already
criticised \cite{Bark2} the theoretical foundation of the model of Kim
et al.

Langanke and Shoppa used in their estimate of the E2 cross section
the model of Kim et al. \cite{Kim} which
clearly appears to be at the high end
of the calculated E2 cross section, or the B(E2) of the 632
keV resonance. While
LS do not list the value they used for $S_{E2}/S_{E1}$ we assume (based on
a private communication) that for the 600 keV angular distribution
they originally used the value of
$S_{E2}/S_{E1}\ =\ 1.8\times 10^{-3}$ which was
modified \cite{Lang2} to the value of $1.4\times 10^{-3}$. This value is
for example a factor 4 larger then would be predicted by Typel and Baur
\cite{Typel} after averaging their results over the energy resolution of the
Riken experiment.  Indeed Typel and Baur calculate E2 cross
sections which are approximately a factor of 4 smaller than
predicted by LS \cite{Lang2}.

We conclude that the uncertainty in nuclear models for the predicted
nuclear E2 cross section does not allow for
a meaningful model dependent estimate of the E2
cross section in the Coulomb Dissociation
of \b8.  This conclusion raises serious doubts on the analysis carried
out by Langanke and Shoppa of the Riken data. We also conclude
that these cross sections (E1 and E2) are best extracted
from a fit to the data, even
if the current data \cite{Moto} are of low precission (15-20\%). Such a fit
is expected to be less uncertain than current model
dependent theoretical estimates.

The predicted angular distribution of the E1 and E2 cross section of the
Coulomb Dissociation of \b8 are sufficiently different \cite{Baur}.  For the
Riken data measured at 46.5 MeV/u the E1 cross section is
dominant at approximately $1-2^\circ$, and the E2 at
$4-5^\circ$.  This should allow in principle for an extraction of the E1
and E2 cross section in the Riken data.

We chose to demonstrate this point for the 600 keV angular distribution
measured at Riken, where LS predict the
largest E2 cross section (see Fig. 1 of LS \cite{Lang1}).
Similar conclusions are found for the other (800 and 1,000 keV)
angular distributions analyzed by LS. We have fitted
the measured Coulomb Dissociation cross section with:
$\sigma_{CD}(E1)\ +\ \sigma_{CD}(E2)$, which are linearly proportional
to $S_{E1}$ and $S_{E2}$, respectively, and the S-factors are
treated as free fit parameters. In Fig. 1 we show the
resulting reduced-$\chi^2$ of this fit to the 600 keV angular distribution
of the Riken data \cite{Moto}.

As shown in Fig. 1 the best fit is obtained for
$S_{E1}\ =\ 18\ \pm\ 3\ eV-b$ and
$S_{E2}\ =\ 0\ \pm\ 6\ meV-b$, corresponding to a 90\% confidence upper
limit of $S_{E2}\ <\ 12\ meV-b$, and $S_{E2}/S_{E1}\ <\ 6\ \times\ 10^{-4}$.
The extracted upper limit
is considerably smaller than that used by LS, but is still consistent with
the lower values predicted by the other models (after averaging
over the experimental energy resolution of the Riken experiment).
Our quoted upper limit contradicts the assumptions of LS and do not
substantiate their analysis, but in fact it confirms
the original assumption of the Riken experiment \cite{Moto} that the data
could be analyzed assuming E1 contribution only.

In conclusion we have demonstrated that theoretical uncertainties in the
estimated E2 cross section of the \be7pg do not allow for a meaningful
model dependent estimate of the E2 cross section of the Riken data
on the Coulomb Dissociation of \b8, and a model independent Chi-Square
analysis yields E2 cross sections that are considerably smaller than
assumed by Langanke and Shoppa.  This invalidates the so called
extracted (or corrected) value of $S_{17}(0)$ as quoted by Langanke and
Shoppa.

\newpage
\begin{center}
{\bf FIGURE CAPTIONS}
\end{center}

\begin{tabbing}

\underline{Fig: 1:} \=The reduced $\chi ^2$ obtained from fitting the 600 keV
angular distribution of the Riken \\
                    \> data \cite{Moto} with:
$\sigma_{CD}(E1)\ +\ \sigma_{CD}(E2)$,
as discussed in the text. \\

\end{tabbing}


\begin{references}

\bibitem{Lang1} K. Langanke and T.D.Shoppa; Phys. Rev. {\bf 49}, R1771(1994).

\bibitem{Lang2} K. Langanke and T.D. Shoppa, Erratum submitted to Phys. Rev.
and private communication.

\bibitem{Moto} T. Motobayashi, N. Iwasa, Y. Ando, M. Kurokawa, H. Murakami,
J. Ruan (Gen), S. Shimoura, S. Shirato, N. Inabe, M. Ishihara, T. Kubo,
Y. Watanabe, M.Gai, R.H. France III, K.I. Hahn, Z. Zhao, T. Makamura,
T. Teranishi, Y. Futami, K. Futami, K. Fututaka, Th. Delbar; preprint
Rikkyo RUP 94-2, Yale-40609-1141, to be published.

\bibitem{Bah} John N. Bahcall, C.A,. Barnes, J. Christensen-Dalsgaard,
B.T. Cleveland, S. Degl'innocenti, B.W. Filippone, A. Glasner, R.W. Kavanagh,
S.E. Koonin, K. Lande, K. Langanke, P.D. Parker, M.H. Pinsonneault,
C.R. Proffitt, and T. Shoppa; placed on the World Wide Web Electronic
Bulletin, 3 April, 1994.

\bibitem{Kim} K.H. Kim, M.H. Park, and B.T. Kim; Phys. Rev.
{\bf C35}, 363(1987).

\bibitem{Bark1} F.C. Barker; Aust. J. Phys. {\bf 33}, 177(1980), and
private communication to T. Motobayashi, 1994.

\bibitem{Bark2} F.C. Barker; Phys. Rev. {\bf C37}, 2920(1988).

\bibitem{Baye} P. Descouvemont, and D. Baye; Nucl. Phys. {\bf A567}, 341(1994).

\bibitem{Krauss} H. Krauss, K. Gr\"{u}n, and H. Oberhummer;
Ann. Physik {\bf 2}, 258(1993).

\bibitem{Typel} S. Typel, and G. Baur, 1994, to be published.

\bibitem{Baur} G. Baur, C.A. Bertulani, and H. Rebel, Nucl. Phys.
{\bf A458}, 188 (1986).

\end{references}
\end{document}